\documentclass[aps, prl, floatfix, nopacs, twocolumn,showkeys, preprintnumbers, nofootinbib, superscriptaddress]{revtex4-2}
\usepackage[utf8]{inputenc}
\usepackage[sort&compress]{natbib}
\usepackage{ulem}
\usepackage{overpic}
\usepackage{bm}
\usepackage{times}
\usepackage{amssymb,amsbsy,amsmath,amsfonts}
\usepackage{graphicx}
\usepackage{float}
\usepackage{color}
\usepackage{booktabs}
\usepackage{makecell}
\usepackage{rotating}
\usepackage{srcltx}
\usepackage{slashed}
\usepackage{subfigure}
\usepackage{multirow}
\usepackage{verbatim}
\usepackage{hyperref}
\usepackage{tabularx}

\DeclareUnicodeCharacter{3002}{HEREHEREHERE}

\begin{document}

\title{Probing the three-body force in hadronic systems with specific charge parity  }

\author{Ya-Wen Pan}
\affiliation{School of Physics,  Beihang University, Beijing 102206, China}

\author{Ming-Zhu Liu}
\email[Corresponding author: ]{liumz@lzu.edu.cn}
\affiliation{
Frontiers Science Center for Rare Isotopes, Lanzhou University,
Lanzhou 730000, China}
\affiliation{ School of Nuclear Science and Technology, Lanzhou University, Lanzhou 730000, China}

\author{Li-Sheng Geng}\email[Corresponding author: ]{lisheng.geng@buaa.edu.cn}
\affiliation{Sino-French Carbon Neutrality Research Center, \'Ecole Centrale de P\'ekin/School of General Engineering, Beihang University, Beijing 100191, China}
\affiliation{School of Physics,  Beihang University, Beijing 102206, China}
\affiliation{Peng Huanwu Collaborative Center for Research and Education, Beihang University, Beijing 100191, China}
\affiliation{Beijing Key Laboratory of Advanced Nuclear Materials and Physics, Beihang University, Beijing 102206, China }
\affiliation{Southern Center for Nuclear-Science Theory (SCNT), Institute of Modern Physics, Chinese Academy of Sciences, Huizhou 516000, China}

\begin{abstract}
Three-body forces, a type of non-perturbative strong interaction, are widely studied in nuclear physics. However, whether their inclusion is necessary in nuclear systems remains a topic of intense debate. 
In this letter, we propose that the existence of three-body forces in certain three-body hadronic systems with definite $C$-parity is certain. Such systems consist of two components whose interactions are mediated by three-body forces—a mechanism not easily realized in conventional three-nucleon systems. 
We investigate two specific three-body hadronic systems, $\bar{D}_sDK$ and $\bar{D}^*D\eta$, using contact-range potentials. The two-body hadron-hadron interactions are constrained by reproducing their scattering lengths, while the three-body couplings are constrained by 
charge symmetry.  
Our results indicate that three-body forces play a minor role in binding the $I(J^{PC})=0(0^{--})$ $\bar{D}_sDK$ system, but a crucial one in binding the $I(J^{PC})=0(1^{-+})$ $\bar{D}^*D\eta$ system. In fact, three-body forces determine whether $\bar{D} ^*D\eta$ forms a bound state, making this system a promising candidate for exploring three-body forces in hadronic physics.                                     

\end{abstract}

\maketitle

\section{Introduction}

Governed by Quantum Chromodynamics (QCD), the strong interaction binds quarks and gluons into nucleons, which account for more than 99\% of the visible mass in the universe. A remarkable feature of QCD is its strong coupling at low energies, posing significant challenges for first-principles studies at the quark level. As a result, investigating hadrons provides a valuable pathway to explore strong interactions. A prominent example is the Cornell potential~\cite{Eichten:1974af}, which was developed through studies of the charmonium spectrum and captures two fundamental aspects of the strong interaction in both low- and high-energy regimes. Moreover, recent studies on exotic hadronic states underscore the essential role of hadron-hadron interactions in understanding their properties.  The study of hadron-hadron interactions is widely discussed in nuclear physics, as the nuclear force serves as an essential input for understanding the structure and properties of atomic nuclei.  Despite the high precision of nucleon forces, such as the  Argonne $v_{18}$ (AV18)~\cite{Wiringa:1994wb} potential, Bonn~\cite{Machleidt:2000ge} potential, and chiral effective field theory(EFT) potentials~\cite{Epelbaum:2014sza,Entem:2017gor,Lu:2021gsb}, the binding energies of $^3$H and $^4$He cannot be accurately described by the NN interaction  alone~\cite{Pudliner:1995wk,Navratil:2000gs}. This indicates that the inclusion of an additional type of non-perturbative strong interactions—the three-nucleon (3N) interaction—is necessary.

The phenomenological 3N interaction was first proposed by Fujita and Miyazawa, arising from the virtual excitation of a $\Delta(1232)$ baryon in processes involving three nucleons interacting via two-pion exchanges~\cite{Fujita:1957zz}. 
 The contribution of this 3N force to the binding energies of the triton and nuclear matter was estimated in Refs.~\cite{Loiseau:1967aqb} and~\cite{Loiseau:1971ymc}, respectively. 
This long-range 3N interaction has since been further developed using more refined models~\cite{Coon:1978gr,Coelho:1983woa}. 
Later, the Urbana and Argonne group supplemented the two-pion exchange by a phenomenological short-range part~\cite{Carlson:1983kq}. In these models, the 3N interaction is decoupled from NN scattering, leading to model dependence in their predictions~\cite{Hammer:2012id}.  A more consistent framework, based on chiral effective field theory (EFT), has been developed to unify the treatment of both NN and 3N interactions and to quantify theoretical uncertainties~\cite{vanKolck:1994yi,Friar:1998zt,Epelbaum:2002vt,Epelbaum:2008ga,Machleidt:2011zz,Lu:2025syk}. Within chiral EFT, 3N interactions arise naturally at next-to-next-to-leading order (N$^2$LO), providing a systematic expansion in which the long- and intermediate components of the nuclear interactions are mediated by pion exchanges and the short-range component is encoded by contact terms~\cite{Epelbaum:2008ga}. As a result, it has become common practice to use chiral EFT with consistent NN and 3N interactions to compute physical observables and study three-nucleon interactions~\cite{Hebeler:2020ocj}. To date, 3N interactions are recognized as essential for reliable predictions of light nuclear properties and the behavior of nucleonic matter~\cite{Machleidt:2011zz,Hammer:2012id,Carlson:2014vla,Hammer:2019poc}, at least in the non-relativistic framework.

According to the analysis based on chiral EFT~\cite{vanKolck:1994yi}, the contribution of 3N interactions relative to that of NN interactions is estimated to be about $5\%$, which is consistent with their respective contributions to the binding energies of light nuclei~\cite{Hammer:2012id}.   
This makes it challenging to isolate and study 3N interactions using physical observables. 
 The long-standing ``$A_y$ puzzle"~\cite{Shimizu:1995zz,Gloeckle:1995jg}, a large discrepancy between experimental measurements and theoretical predictions of a particular physical observable in elastic nucleon-deuteron scattering, is sensitive to the value of $c_D$ in the  3N interactions~\cite{LENPIC:2018ewt}.    $c_D$ is a low-energy constant appearing in the chiral EFT description of the three-nucleon force at N$^2$LO, which is determined from observables in three-nucleon systems~\cite{LENPIC:2018ewt}. 
Moreover, the $P$-wave phase shifts in neutron–$\alpha$ scattering are also sensitive to 3N interactions~\cite{Lynn:2015jua}. Recently, Yang et al. suggested that the $D$-wave phase shifts in neutron–$\alpha$ scattering provide a favorable probe of the long-range behavior of 3N interactions~\cite{Yang:2025mhg}.
Additionally, including 3N interactions significantly improves the description of nuclear properties, such as spectra~\cite{Soma:2019bso} and charge radii~\cite{Hoppe:2019uyw}.
Nevertheless, 3N interactions remain less well understood than two-body NN interactions. For example, it has recently been shown that the saturation of nuclear matter can be reasonably described without explicitly including 3N interactions in the Dirac-Brueckner-Hartree-Fock approach with leading order and next-to-leading order covariant chiral nuclear forces~\cite{Zou:2023quo,Zou:2025dao}. In this Letter, we propose investigating three-body forces by replacing three nucleons with three hadrons of specific charge parities.

In a three-nucleon system, there is no well-defined charge-parity. However, when nucleons are replaced by mesons — such as in the $\bar{D}_sDK$ system — the resulting three-meson combination can possess a definite charge parity, implying that the system is composed of two distinct components: $\bar{D}_sDK$ and ${D}_s\bar{D}\bar{K}$. In the Hamiltonian of this system, a term appears that describes the transition from the $\bar{D}_sDK$ component to the ${D}_s\bar{D}\bar{K}$ component, and it depends on the charge-parity~\cite{Wu:2025fzx}. Notably, this transition corresponds to a three-body force, indicating that a three-body force naturally arises in such a system with specific charge parity, similar to the 3N potentials in Chiral EFT~\cite{Epelbaum:2008ga}. This offers a unique opportunity to study three-body forces directly.

In this letter, we use pionless EFT to investigate three-body hadronic systems with definite charge-parity, i.e.,  $I(J^{PC})=0(0^{--})$ $\bar{D}_sDK$ and $I(J^{PC})=0(1^{-+})$ $\bar{D}^*D\eta$. These three-body systems carry the quantum number beyond conventional charmonium states, whose non-perturbative effect, such as the bare $c\bar{c}$ dressed effect, would be suppressed,  making them promising candidates for three-body hadronic molecules~\cite{Liu:2024uxn}.    We find that the three-body force plays a more prominent role in binding the $\bar{D}^*D\eta$ system than in binding the $\bar{D}_sDK$ system, making $\bar{D}^*D\eta$  a promising candidate for probing three-body interactions.

\section{Theoretical Framework}

\begin{figure}[!htbp]
  \centering
  \begin{overpic}[scale=0.4]{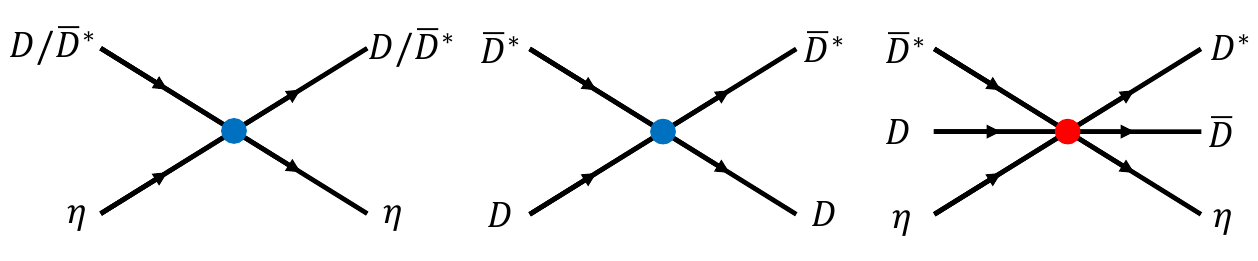}
  \end{overpic}
  \caption{Contact term in the $\bar{D}^*D\eta$ system.}
  \label{3f}
\end{figure}

\begin{figure}[ttt]
  \centering
  \begin{overpic}[scale=0.63]{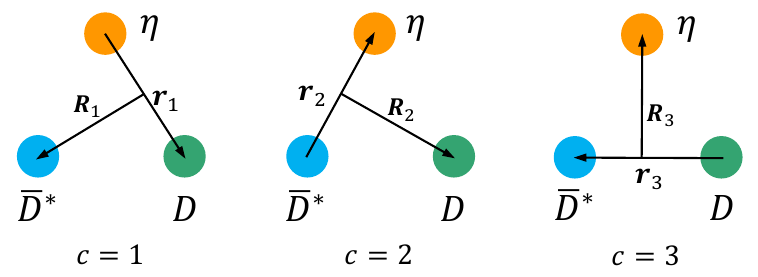}
  \end{overpic}
  \caption{ Three Jacobi channels of the $\bar{D}^*D\eta$ system.}
  \label{jaco}
\end{figure}

In this letter, we investigate the $\bar{D}_sDK$ and $\bar{D}^*D\eta$ systems with definite charge parity. The wave functions of these systems consist of two components: $\bar{D}_sDK$ ($D_s\bar{D}\bar{K}$) and $\bar{D}^*D\eta$ (${D}^*\bar{D}\eta$).
We study these three-body systems by solving the Schr\"{o}dinger equation with the Gaussian expansion method (GEM)~\cite{Hiyama:2003cu}.
\begin{equation}
   \left[T + \sum_{1 = i < j}^{3} V(r_{ij}) + V^C - E \right] \Psi_{JM}^{\mathrm{Total}} = 0\,,
\end{equation}
where $T$ denotes the kinetic energy and $V(r_{ij})$ represents the two-body potential corresponding to the coordinate $r_{ij}$.   $V^C$ stands for the potential dependent on the charge parity, such as the process $\bar{D}^*D\eta \to D^*\bar{D}\eta$ depicted in Fig.~\ref{3f}, which corresponds to a three-body potential, like the contact term of 3N potentials in Chiral EFT~\cite{Hebeler:2020ocj}.  The total wave function, $\Psi_{JM}^{\mathrm{Total}}$, is expressed as a sum of the wave functions of three Jacobi channels, as illustrated in Fig.~\ref{jaco}, 
\begin{equation}
    \Psi_{JM}^{Total} = \sum_{c=1}^{3} \Phi_{JM,\alpha}^{c}(\pmb{r}_{c},\pmb{R}_{c})\,.
\end{equation}
The wave function of each channel  $\Phi_{JM,\alpha}^{c}(\pmb{r}_{c},\pmb{R}_{c})$ is expanded in terms of a finite set of localized, rapidly decaying Gaussian basis functions:
\begin{equation}
    \Phi_{JM,\alpha}^{c}(\pmb{r}_{c},\pmb{R}_{c}) = \sum_{\alpha}C_{c,\alpha} \left[\phi^{G}_{n_{c}l_{c}}(\pmb{r}_{c})\psi^{G}_{N_{c}L_{c}}(\pmb{R}_{c})\right]_{JM}\otimes H_{t}^{c}\,,
\end{equation}
with 
\begin{equation}
    \begin{aligned}
    &\phi^{G}_{nlm}(\pmb{r})=N_{nl} r^{l} e^{-\nu_{n}r^{2}}Y_{lm}(\hat{\pmb{r}})\,, \\
    &\psi^{G}_{NLM'}(\pmb{R})=N_{NL} R^{L} e^{-\lambda_{N}R^{2}}Y_{LM'}(\hat{\pmb{R}})\,,
    \end{aligned}
\end{equation}
where $C_{c,\alpha}$ are the expansion coefficients. The term $H_t^c = [\chi^i \otimes \chi^j]_t\otimes\chi^k$ represents the total isospin wave function for the Jacobi channel $c$, where $\chi$ denotes the isospin wave function of each particle. Since the total isospin of the $\bar{D}_{s}DK(\bar{D}^*D\eta)$ three-body systems is zero, and the $\bar{D}_s(\eta)$ mesons themselves carry isospin zero, it follows that the $DK(\bar{D}^*D)$ two-body subsystems must also have isospin zero.
The index $\alpha$ corresponds to a set of labels $\{n, N, l, L, t\}$, where $l$ and $t$ are the orbital angular momentum and isospin for the Jacobi coordinate $\pmb{r}$, and $L$ is the orbital angular momentum for the Jacobi coordinate $\pmb{R}$. Here, we focus only on $S$-wave interactions. For further details on the GEM, see Refs.~\cite{Pan:2022xxz, Pan:2024ple, Pan:2025xvq}.

For the $\bar{D}_sDK$ and $\bar{D}^*D\eta$ systems, two-body potentials exist between the hadron pairs  $DK$, $\bar{D}_sK$, $\bar{D}_sD$, $\bar{D}^*D$, $\bar{D}^*\eta$, and $D\eta$. These potentials are constructed and determined directly in the isospin basis; therefore, the explicit particle charges are not shown. In this letter, we adopt contact-range potentials within the EFT framework as shown in Fig.~\ref{3f}, which take a Gaussian form in coordinate space:
\begin{equation}
    V = C_a e^{-r^2/b_2^2}\,,
    \label{contactrange1}
\end{equation}
where $C_a$ denotes the strength of the corresponding two-body potential and $b_2$ is the cutoff. In this work, we consider values of $b_2$ ranging from $0.5$  to $1.0$ fm~\cite{Liu:2018zzu}.   As indicated in Refs.~\cite{Kaplan:1998tg,Kaplan:1998we,Fleming:2007rp,Nieves:2012tt,Du:2021zzh}, the one-pion exchange(OPE) term is expected to be of higher order and is partially absorbed into the contact term via renormalization. In addition, we have verified that the OPE potential has a negligible effect on the three-body system. Therefore,  only  the contact-range potential is considered in this letter.

Next, we explain how to determine the values of $C_a$ by reproducing the corresponding scattering data. For a general two-body scattering process, the $S$-wave phase shift is given by $\delta_0(k) = \mathrm{arg}(S(E'))$, where $S(E')$ is the scattering matrix. The phase shift can be further parameterized in terms of the scattering length $a_0$ and effective range $r_0$~\cite{Luscher:1990ux}:
\begin{equation}\label{ar}
  k\mathrm{cot}\delta_0(k)=\frac{1}{a_0}+\frac{1}{2}r_0 k^2+\mathcal{O}(k^4)\,.
\end{equation}
This shows that $a_0$ and $r_0$ are related to the scattering matrix $S(E')$. 
In general, the $S(E')$ is 
determined by the asymptotic behavior of the two-body wave function $u(r)$: 
\begin{equation}
    u(r)\overset{r\to\infty}{\longrightarrow} e^{-ikr} -S(E')e^{ikr}\,,
\end{equation}
where the momentum is defined as $k(E') = \sqrt{2\mu(E'-M)}$,  with  $\mu$ being the reduces mass of the two-body system   and $M$ the corresponding mass threshold. The wave function $u(r)$ is obtained by solving the two-body Schr\"{o}dinger equation $(T+V-E')u(r)=0$ using the Numerov method~\cite{Yalikun:2021bfm}. Here, the potential term $V$ corresponds to $C_a$. Therefore, using the values of $a_0$ and $r_0$, one can extract the values of $C_a$. In this letter, we determine $C_a$ using the scattering lengths $a_0$ and the effective range $r_0$ obtained from lattice QCD simulations or invariant-mass distributions.  For systems for which reliable scattering lengths are unavailable, we estimate them from symmetry arguments.

For $DK$ scattering, 
the lattice QCD simulations  reported $a_0 = -1.49\pm 0.13$ fm and $r_0 = 0.20\pm 0.09$ fm at $m_{\pi} = 150$ MeV~\cite{Bali:2017pdv}, where the dressed effect of bare $c\bar{s}$ is taken into account. 
Therefore, we also consider the potential describing the $c\bar{s}$ bare state coupling to $DK$, characterized as $V' = C_{c\bar{s}}e^{-r^2/b_{c\bar{s}}^2}$~\cite{Yamaguchi:2019vea}. Because the range of the interaction term $V^{\prime}$ is smaller than that of $V$ according to the meson-exchange theory, we set $b_{c\bar{s}}=0.4$ fm. The values of the $DK$ potential, $C_{a}$, are determined as in Table~\ref{CDDbs}.  One should note that changing $b_{c\bar{s}}$ has a negligible effect on $C_a$.  With the $DK$ potential in Table~\ref{CDDbs}, we estimate   the $c\bar{s}$  composition in $D_{s0}^*(2317)$, i.e.,  $\mathcal{P}(c\bar{s}) = 21\% \sim 38\%$,  consistent with Refs.~\cite{MartinezTorres:2014kpc,Albaladejo:2018mhb,Yang:2021tvc,Guo:2023wkv,Gil-Dominguez:2023huq}. For the $\bar{D}_sK$ potential, we adopt the relation derived in Ref.~\cite{Wu:2025fzx}. i.e., $V_{\bar{D}_sK} = \frac{1}{2}V_{DK}$. 
The scattering lengths of the  $\bar{D}_sD$, $\bar{D}^*D$, and $D\eta$ systems  are $a_{\bar{D}_sD} = 0.525^{+0.07}_{-0.06}$ fm~\cite{Ji:2022vdj}, $a_{\bar{D}^*D} = -1.7\pm 0.4$ fm~\cite{Prelovsek:2013cra}, and $a_{D\eta}=0.29^{+0.15}_{-0.22}$ fm~\cite{Guo:2018tjx}.   The scattering length for $\bar{D}_sD$ is extracted from the low energy constant $\mathcal{C}_{1a}$ in Ref.~\cite{Ji:2022vdj} assuming SU(3) flavor symmetry. 
 Within uncertainties, the scattering length $a_{D\eta}$ adopted in this work agrees with the value of about 0.38~fm extracted from the lattice QCD data reported by the Hadron Spectrum Collaboration~\cite{Moir:2016srx}.
 Note that the potentials determined by simultaneously reproducing the scattering length and, when available, the effective range lead to the same conclusion. To ensure consistency across all potentials, we set the cutoffs for the $\bar{D}_sD$, $\bar{D}^*D$, and $D\eta$ potentials equal to those for the $DK$ potential, and determine the potential strengths by fitting the corresponding scattering lengths.
By fitting these scattering lengths, the values of the corresponding potential $C_a$ are collected in Table~\ref {CDDbs}.  The $D^*\eta$ potential is determined by the $D\eta$ potential  via the heavy quark spin symmetry.   Since the strong interaction is invariant under charge conjugation, the potentials for antiparticle-antiparticle systems (such as $\bar{D}\bar{K}$, $D_s\bar{K}$, $D_s\bar{D}$, $\bar{D}^*D$, $\bar{D}\eta$, and $\bar{D}^*\eta$) are identical to those of the corresponding particle-particle systems determined above.

\begin{table}[!h]
\setlength{\tabcolsep}{4pt}
\centering
\caption{Strengths of the $DK$, $\bar{D}_sD$, $\bar{D}^*D$, and $D\eta$ potentials $C_{DK}$, $C_{\bar{D}_sD}$, $C_{\bar{D}^*D}$, and $C_{D\eta}$ (in units of MeV) determined by fitting the corresponding scattering lengths and effective ranges. \label{CDDbs}}
\begin{tabular}{ccccccc}
  \hline\hline
   Cutoff & $C_{DK}$  &$C_{\bar{D}_sD}$ & $C_{\bar{D}^*D}$ & $C_{D\eta}$\\
  \hline
     $b_2 = 0.5$ fm  & $-749^{+53}_{-64}$ & $-104^{+7}_{-7}$ & $-316^{+25}_{-49}$ & $-165^{+110}_{-48}$\\
     $b_2 = 1.0$ fm  & $-284^{+41}_{-55}$ & $-16.9^{+1.4}_{-1.5}$ & $-128^{+22}_{-38}$ & $-24.4^{+17.5}_{-9.3}$\\
 \hline \hline
\end{tabular}
\end{table}

Similar to the Gaussian-type potential in a two-body system, the contact term of the three-body potential in coordinate space is expressed as~\cite{Machleidt:2024bwl}:
\begin{equation}
    V^C = C_3e^{-r^2/b_3^2}e^{-R^2/b_3^2},
\end{equation}
where the parameter $C_3$ characterizes the strength of the three-body force, and the parameter $b_3$ represents the cutoff. 
Since the properties of the three-body potential in hadronic systems are still poorly understood, we take $b_3=0.3$ and $0.5$ fm in this work, assuming that the interaction range of the three-body potential is shorter than that of the two-body potential. The strength $C_3$ is treated as a free parameter, and the charge parity determines its sign. Following Ref.~\cite{Wu:2025fzx}, the $C_3$  term  for the $I(J^{PC})=0(0^{--})$ $\bar{D}_sDK$ system is repulsive.  Since the $I(J^{PC})=0(1^{++})$ $\bar{D}^*D$ two-body potential dependent on the charge parity is attractive~\cite{Sun:2011uh}, it can be expected that the $C_3$ term for  $I(J^{PC})=0(1^{-+})$ $\bar{D}^*D\eta$ is attractive as well.

\section{results and discussions}

It is important to note that the three-body system treated in the GEM is effectively confined to a finite spatial volume because a finite, localized basis set is used. As a result, continuum states, especially those describing the free motion of one particle, cannot be accurately described. Instead, such states appear as discrete eigenvalues, forming a discretized continuum that can sometimes mimic bound states. Therefore, a three-body bound state can only be confidently claimed if the lowest eigen energy $E_{\mathrm{min}}<0$ and the corresponding root-mean-square(rms) radii $\left\langle r \right\rangle$, $\left\langle R \right\rangle$ are both finite and stable under the basis variation. This criterion is particularly critical when a repulsive three-body potential is included, as discussed in Ref.~\cite{Pan:2025xvq}.

\begin{figure}[ttt]
  \centering
  \begin{overpic}[scale=0.435]{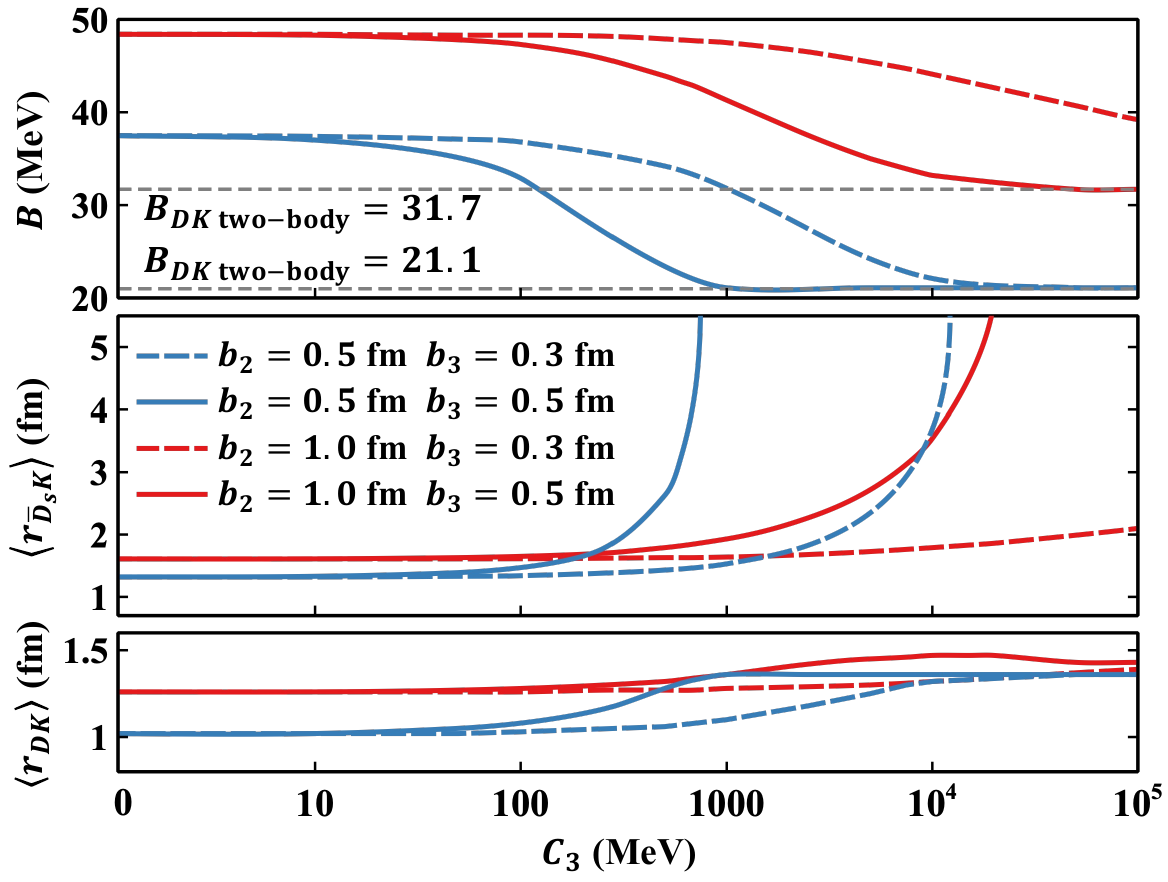}
  \end{overpic}
  \caption{Binding energy and rms radii $\langle r_{\bar{D}_s K} \rangle$ and $\langle r_{D K} \rangle$ of the $I(J^{PC})=0(0^{--})$~$\bar{D}_sDK$ three-body system as functions of the three-body potential strength $C_3$, shown for different choices of the cutoffs $b_2$ and $b_3$.  \label{pic_1}}
\end{figure}

There exist two unknown cutoff parameters,  $b_2$ and $b_3$,  for the two-body interactions and three-body interactions, respectively.   The values of the two-body interactions $C_a$  are determined at the cutoff of $b_2 = 0.5$ and $1.0$ fm, as listed in Table~\ref{CDDbs}. For the three-body interaction, the cutoff is set to $b_3=0.3$ or $0.5$ fm, and the values of $C_3$ are treated as a free parameter. In this letter, we consider four different combinations of $b_2$ and $b_3$. 
 We have verified that the uncertainties in the potentials have a negligible effect on the $\bar{D}_sDK$ system, and, therefore, we focus on their central values for clarity in the following analysis of the $\bar{D}_sDK$ system. 
In Fig.~\ref{pic_1}, we present the binding energies and rms radii of the $I(J^{PC})=0(0^{--})$~$\bar{D}_sDK$ three-body system as functions of the three-body potential strength $C_3$ for these four cases. Our results indicate that as the strength of the three-body potential increases, the binding energy of the three-body system decreases and approaches that of the $DK$ two-body system, as shown in the top panel of Fig.~\ref{pic_1}. Meanwhile, the rms radius $\langle r_{DK}\rangle$ increases gradually from about $1$ to $1.5$~fm, while $\langle r_{\bar{D}_s K}\rangle$ expands until it eventually reaches the boundary of the computational volume, as shown in the middle and bottom panels of Fig.~\ref{pic_1}. These behaviors suggest that, in the limit of a strong three-body interaction, the $\bar{D}_sDK$  molecule can be viewed as a $\bar{D}_s$ meson and a two-body $DK$ molecule.  It should be noted that the rms radius of the two-body subsystem is determined not solely by its potential, but also by the  potentials of other subsystems  as well as  the three-body potential. 
Furthermore, a larger $b_3/b_2$ ratio amplifies the influence of the three-body potential on the system. Although a strongly repulsive three-body potential could, in principle, break the $\bar{D}_sDK$ bound state, this occurs only when $C_3 > 1000$~MeV—a scenario unlikely under typical conditions. 
For $C_3 < 1000$ MeV, the expectation value of the three-body potential accounts for at most $4\%$ of the total potential energy, as estimated from $\frac{|\langle V_3\rangle|}{|\langle V_3\rangle| + \sum_{1=i<j}^{3}|\langle V(r_{ij})\rangle|}$.  
Thus, we conclude that the repulsive three-body force has only a minor effect on the $I(J^{PC})=0(0^{--})$~$\bar{D}_sDK$ bound state, in agreement with the three-body force characterized by two-body subsystem potentials in Ref.~\cite{Wu:2025fzx}. Therefore, we turn to the $I(J^{PC})=0(1^{-+})$ $\bar{D}^*D\eta$ system to explore the three-body force.

\begin{figure}[ttt]
  \centering
  \begin{overpic}[scale=0.445]{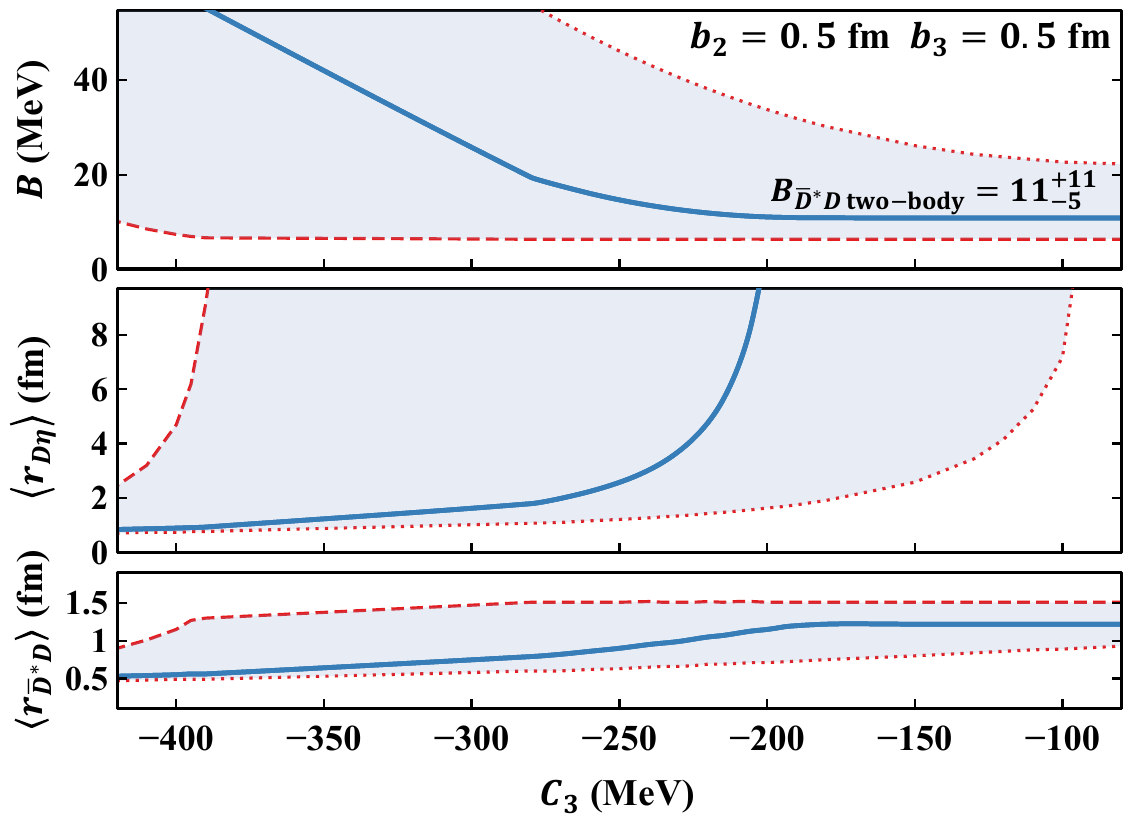}
  \end{overpic}
  \caption{Binding energy and rms radii $\langle r_{D\eta} \rangle$ and $\langle r_{\bar{D}^*D} \rangle$ of the $I(J^{PC})=0(1^{-+})$~$\bar{D}^*D\eta$ three-body system as functions of the three-body potential strength $C_3$ for $b_2=0.5$ fm.  The blue solid lines represent the results obtained from the central values of the two-body potentials, and the red dotted (dashed) lines represent the extreme cases where all two-body potential strengths are the strongest (weakest). } \label{pic_2}
\end{figure}

\begin{figure}[!h]
  \centering
  \begin{overpic}[scale=0.445]{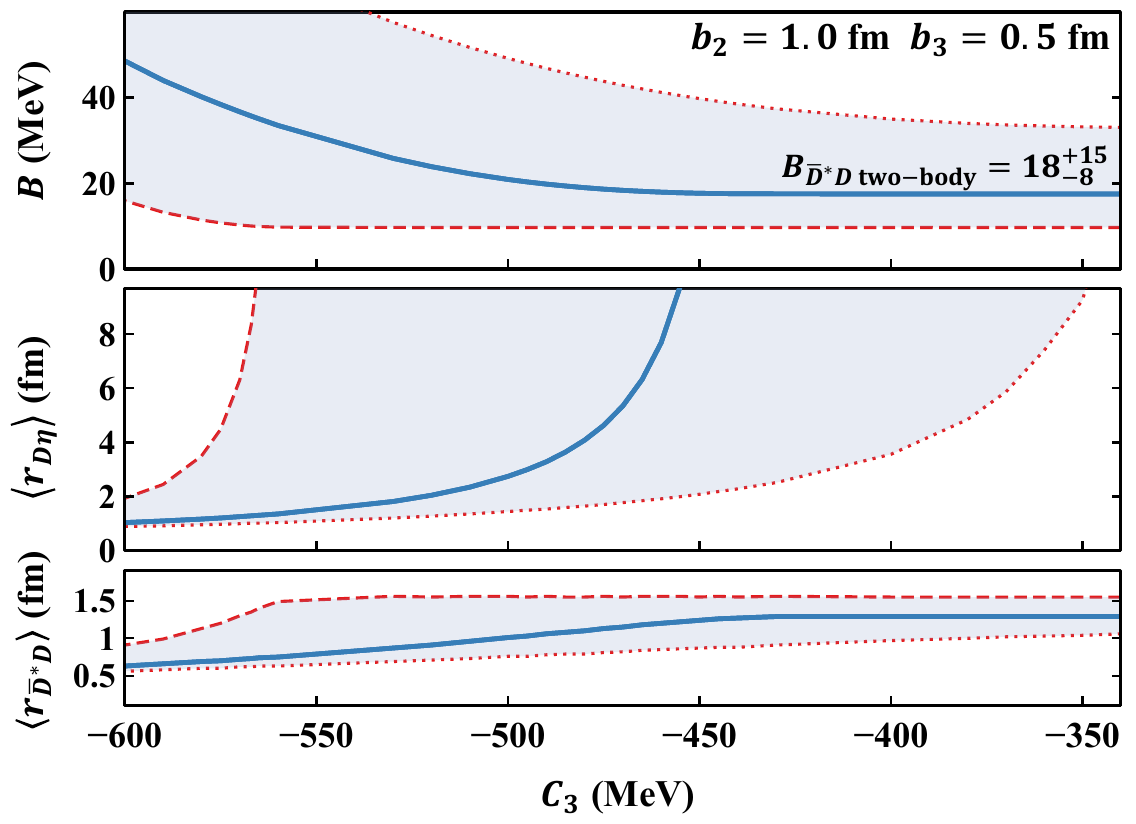}
  \end{overpic}
  \caption{Same as Fig.~\ref{pic_2}   but for $b_2=1.0$ fm. \label{pic_3}}
\end{figure}

Since the results for the  $I(J^{PC})=0(0^{--})$~$\bar{D}_sDK$ system are not affected in any significant way by changes in the cutoff $b_3$, we fix $b_3 = 0.5~\text{fm}$ in the following analysis of the $I(J^{PC})=0(1^{-+})$ $\bar{D}^*D\eta$ system.  
In Fig.~\ref{pic_2}, we  show the  binding energy  and rms radii $\langle r_{D\eta} \rangle$ and $\langle r_{\bar{D}^*D} \rangle$  with the two-body cutoff set to $b_2 = 0.5~\text{fm}$. Additionally, we have verified that $\langle r_{D\eta} \rangle \approx \langle r_{\bar{D}^*\eta} \rangle$ for $\langle r_{\bar{D}^*\eta} \rangle$. 
 Considering two extreme cases within the range of potential uncertainties, in which all two-body potential strengths are the strongest (red dotted lines) and weakest (red dashed lines), the results remain consistent. Therefore, we focus on the central values (blue solid lines) in the subsequent numerical analysis for clarity. 
Although solving the three-body Schr\"{o}dinger equation without a three-body potential yields a binding energy of $10.9$ MeV, the corresponding rms radii $\langle r_{D\eta} \rangle$ and $\langle r_{\bar{D}^*\eta} \rangle$ consistently extend to the boundary of the computational domain. This indicates the absence of a $\bar{D}^*D\eta$ three-body bound state in this scenario.  As the three-body potential becomes more attractive, a $I(J^{PC})=0(1^{-+})$~$\bar{D}^*D\eta$ three-body bound state begins to form when the strength of the three-body potential reaches about $C_3 \simeq -195$ MeV, which implies the importance of the three-body force in forming such a bound state.  This is clearly illustrated in the top and middle panels of Fig.~\ref{pic_2}: for $C_3 < -195$ MeV, the binding energy increases gradually, whereas $\langle r_{D\eta} \rangle$ becomes finite and stable across variations in the basis. With $C_3 = -240$ MeV, we obtain a $I(J^{PC})=0(1^{-+})$~$\bar{D}^*D\eta$ three-body bound state with a binding energy of approximately $14$ MeV and rms radius in the range of $1 \sim 3$ fm, consistent with the typical size of a hadronic molecule. A similar conclusion holds for a larger two-body cutoff $b_2=1.0$ fm, where a $\bar{D}^*D\eta$ three-body bound state emerges when the strength of the three-body potential reaches about $C_3 \simeq -450$ MeV, as shown in  Fig.~\ref{pic_3}. The absolute value of $C_3$ for the $I(J^{PC})=0(1^{-+})$ $\bar{D}^*D\eta$ system falls within the range analyzed in Fig.~\ref{pic_1}, suggesting that the existence of such a molecular state is likely.

More importantly, even if the $\bar{D}^*D$ potential is weakened to the point where the $\bar{D}^*D$ subsystem forms only a shallow bound state with a binding energy of $4$ MeV, and even in the absence of any interaction between the $D(\bar{D}^*)$ and $\eta$ mesons, a three-body bound state can still emerge once the three-body potential becomes sufficiently attractive.   For instance, with $C_3 = - 490 $ MeV and $b_2=b_3=0.5$ fm, a $\bar{D}^*D\eta$ bound state appears with a binding energy of about $5.3$ MeV and rms radius of $1\sim 4$ fm. Moreover, given the presence of a $\bar{D}^*D\eta$ bound state, the expectation value of the three-body potential accounts for at least $18\%$ of the total potential energy from the analysis of Fig.~\ref{pic_2} and Fig~\ref{pic_3}. However, this ratio is around $4\%$ for the $I(J^{PC})=0(0^{--})$~$\bar{D}_sDK$ system at most.      The above analysis demonstrates that the three-body interaction plays a significant role in binding the $\bar{D}^*D\eta$ system, suggesting that the $\bar{D}^*D\eta$ system is a promising candidate for probing three-body interactions in hadronic systems.

 Following Ref.~\cite{Wu:2025fzx}, we employ an effective Lagrangian approach to compute the strong decays of the $I(J^{PC})=0(1^{-+})$ $\bar{D}^*D\eta$ molecule(denoted by $X(4412)$) and its production in $B$ meson decays.  In this framework, the $X(4412)$ is generated through the $X(3872)\eta$ channel, given that the $\bar{D}^*D$-$\eta$ subsystem, corresponding to Fig.~\ref{jaco} (c), accounts for more than $80\%$ of its wave function. Further details are provided in the Supplemental Material. Our calculations yield the following branching fractions:  $\mathcal{B}[B \to  (X(4412) \to D^{*\pm}D^{\mp}) K] \approx 0.1\times 10^{-6}$, $\mathcal{B}[B \to   (X(4412) \to D^+D^- ) K] \approx 0.5\times 10^{-6}$,  and $\mathcal{B}[B \to   (X(4412) \to D^{*+}D^{*-})K] \approx 0.4\times 10^{-6}$. With the experimental branching fractions  $\mathcal{B}(B \to D^{*\pm}D^{\mp} K)=6\times 10^{-4} $,  $\mathcal{B}(B \to D^{+}D^{-} K) = 2.2 \times 10^{-4}$, and $\mathcal{B}(B \to D^{*+}D^{*-} K) = 1.32 \times 10^{-3}$~\cite{ParticleDataGroup:2024cfk}, the corresponding  ratio    $\mathcal{B}[B \to  (X(4412) \to D^{(*)+}D^{(*)-} )K]/\mathcal{B}(B \to D^{(*)+}D^{(*)-} K)$ are estimated to be around $2\times 10^{-4}$,  $2\times 10^{-3}$, and $3\times 10^{-4}$. Finally, with the large event number for the decays $B \to D^{*\pm}D^{\mp} K$ and $B \to D^{+}D^{-} K$ being $2\times 10^3$~\cite{LHCb:2024vfz} and $1\times 10^3$~\cite{LHCb:2020bls} at luminosities of $9~\text{fb}^{-1}$ from the LHCb Collaboration,   we obtain the corresponding event numbers of observing $X(4412)$  in $B$ decays as $0.4$ and $2$,  indicating that it is difficult to be observed in current experiments.  The event numbers of observing $X(4412)$ in the decays $B \to D^{*\pm}D^{\mp} K$ and $B \to D^{+}D^{-} K$  would increase to a couple of tens for luminosities of $50~\text{fb}^{-1}$ and $350~\text{fb}^{-1}$. The event number for the decay $B \to D^{*+}D^{*-} K$ from the BaBar Collaboration is approximately $2\times 10^2$~\cite{BaBar:2010tqo}, implying that the expected number of $X(4412)$ events is only $0.06$, which renders its detection in this channel highly challenging.  Therefore,  experimental searches for the $I(J^{PC})=0(1^{-+})$ $\bar{D}^*D\eta$ molecule in the channels $B \to D^{*\pm} D^{\mp} K$ and $B \to D^+ D^- K$ are promising at future LHCb experiments.

\section{Summary and outlook}
\label{sum}

In this letter, we investigate the three-body force in three-hadron systems with definite charge parity, where such forces naturally manifest themselves through the dependence of the three-body interaction on the charge parity. We adopt contact-range potentials from EFT to describe both two- and three-body interactions. The low-energy constants in the two-body sector are constrained by fitting to scattering lengths, while those associated with three-body interactions are constrained by charge parity. The criteria for the formation of a three-body molecular state are its binding energy and the rms radii of its subsystems.   Our findings suggest that the three-body force plays only a minor role in binding the $I(J^{PC})=0(0^{--})$~$\bar{D}_sDK$ system, whereas it significantly contributes to the binding of the $I(J^{PC})=0(1^{-+})$~$\bar{D}^*D\eta$ system.   This indicates that the latter $\bar{D}^*D\eta$ three-body molecular state is a promising candidate for probing three-body forces.  We recommend experimental searches for the $I(J^{PC})=0(1^{-+})$ $\bar{D}^*D\eta$ molecule in the channels $B \to D^{*\pm} D^{\mp} K$ and $B \to D^+ D^- K$ at future LHCb experiments.  Beyond the masses, other observables associated with these three-body scattering processes, such as the lineshape of momentum correlation functions~\cite{Kievsky:2023maf,ALICE:2025aur} and invariant-mass distributions~\cite{Lin:2025ksg}, can also provide valuable insights into the nature of three-body forces.

\section{Acknowledgments}
 
 This work is partly supported by the National Key R\&D Program of China under Grant No. 2023YFA1606703 and the National Natural Science Foundation of China under Grant No. W2543006 and No. 12435007. M.Z.L acknowledges support from the National Natural Science Foundation of China under Grant No.~12575086. Y.W.P acknowledges support from the National Natural Science Foundation of China under Grant No. 12547153 and 
 the China Postdoctoral Science Foundation under Grant Number 2025M784262.

\bibliography{sample.bib}

\clearpage

\begin{widetext}
\setcounter{page}{1}
\setcounter{figure}{0}  
\setcounter{table}{0}  
\section{Supplemental Material}

In this Supplemental Material, we provide details on the effective Lagrangian calculation of the strong decays and production of the $I(J^{PC})=0(1^{-+})~\bar{D}^*D\eta$ molecule.

\section{Search for $I(J^{PC})=0(1^{-+})~\bar{D}^*D\eta$ molecule }

Following Ref.~\cite{Wu:2025fzx}, we propose that the decay of the three-body $\bar{D}^*D\eta$ molecule (denoted by $X(4412)$) proceeds via the $\bar{D}^*D-\eta$ subsystem, since the $\bar{D}^*D$-$\eta$ configuration accounts for more than $80\%$ of its wave function. Therefore,  the $X(4412)$ decays primarily through the intermediate two-body subsystem $X(3872)\eta$, and then subsequently transitions into $\bar{D}^*D^*$, $\bar{D}D$, and $\bar{D}D^*/\bar{D}^*D$, as illustrated in Fig.~\ref{strongdecay}(a,b), Fig.~\ref{strongdecay}(c,d), and Fig.~\ref{strongdecay}(e,f), respectively.

\begin{figure}[!h]
	\centering
	\includegraphics[width=12.0cm]{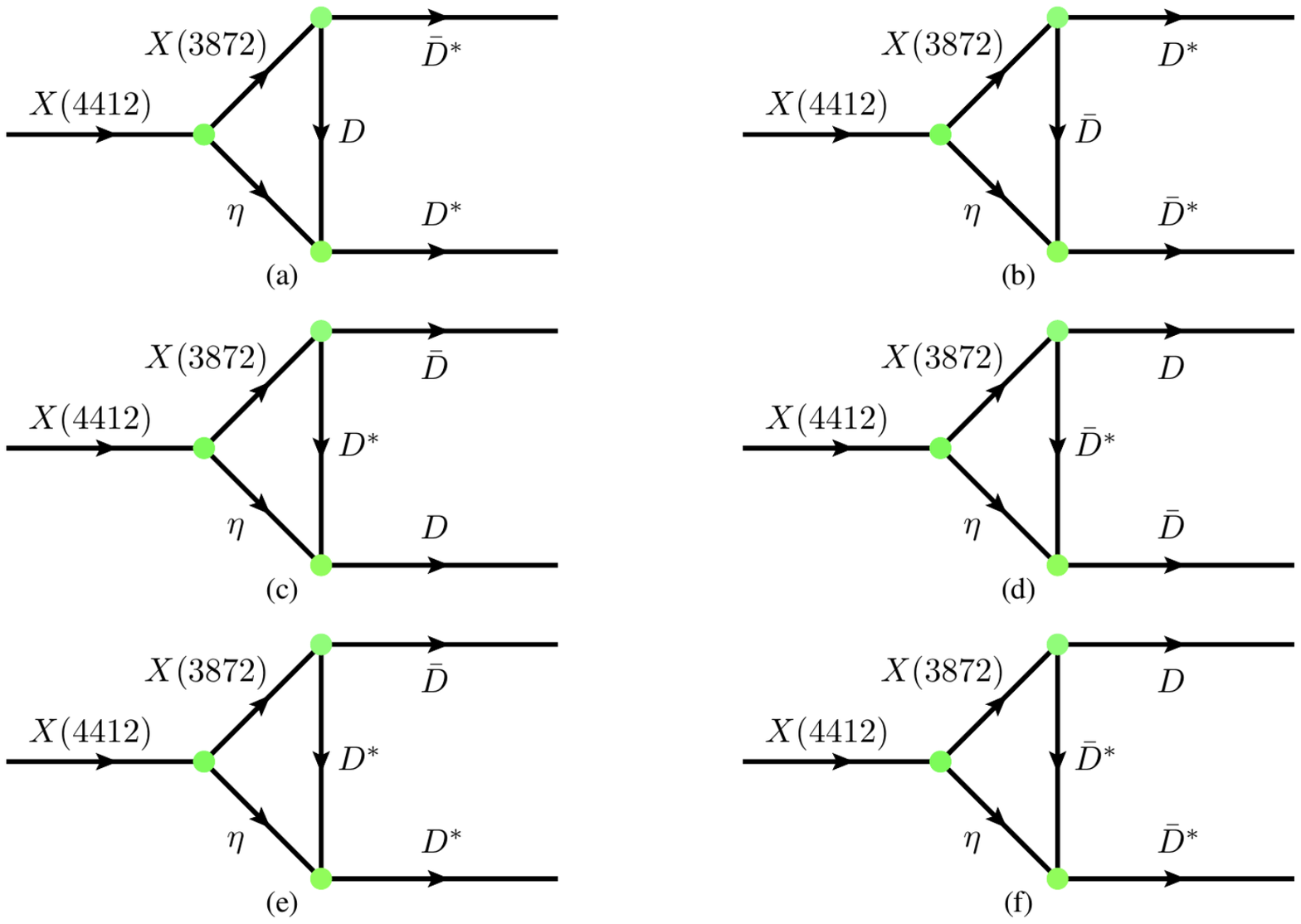}
	\caption{Triangle diagrams for the strong decays of $X(4412) \to \bar{D}^*D^*$(a,b), $X(4412) \to \bar{D}D$(c,d), and $X(4412) \to \bar{D}D^*/D\bar{D}^*$(e,f) .   }\label{strongdecay}
\end{figure}

In this letter, we calculate the partial decay widths for $X(4412) \to \bar{D}^*D^*$, $X(4412) \to \bar{D}D$, and $X(4412)\to \bar{D}^*D/D^*\bar{D}$ through the effective Lagrangian approach.
The effective Lagrangian describing the interaction vertices of Fig.~\ref{strongdecay} are~\cite{Oh:2000qr,Guo:2013zbw,Dong:2022cuw}
\begin{eqnarray}
\label{dddds}
\mathcal{L}_{D^{\ast} D \eta}&=& -i g_{D^{\ast} D \eta} ( D \partial^{\mu} \eta \bar{D}_{\mu}^{\ast}-  D_{\mu}^{\ast}\partial^{\mu} \eta \bar{D}), \\  
\mathcal{L}_{D^{\ast} D^{\ast} \eta}&=& - g_{D^{\ast} D^{\ast} \eta} \varepsilon_{\mu\nu\alpha\beta} \partial^{\mu}D^{\ast\nu} {\partial}^{\alpha} \bar{D}^{\ast\beta} \eta,  \\ 
\mathcal{L}_{X(3872)D\bar{D}^*}&=&  g_{X(3872)D\bar{D}^*} X^{\mu}(3872)(D^{*}_{\mu}\bar{D}-D\bar{D}^*_{\mu}),  \\ 
\mathcal{L}_{X(4412)X(3872)\eta}&=&  g_{X(4412)X(3872)\eta} X^{\mu}(4412)X_{\mu}(3872)\eta, 
\end{eqnarray}
where the couplings are taken as $g_{D^{\ast} D \eta}=11.94/\sqrt{3}$, 
$g_{D^{\ast} D^{\ast} \eta}=g_{D^{\ast} D\eta}/m_{\bar{D}}$, $g_{X(3872)D\bar{D}^*}=9.16/\sqrt{2}$~GeV, and $g_{X(4412)X(3872)\eta}=14.35$~GeV~\cite{Zhang:2025rlg,Liu:2024ziu}. The $m_{\bar{D}}$ represents the average mass of $D$ and $D^*$. 

With the above Lagrangians, the amplitudes of Fig.~\ref{strongdecay} are written as 
\begin{eqnarray}
 \label{gamma}   
 i\mathcal{M}^{a,b}&=&g_{X(4412)X(3872)\eta} g_{X(3872)D\bar{D}^*}g_{D^*D\eta} \int \frac{d^{4}q}{(2\pi)^4} \frac{-g^{\mu\alpha}+\frac{q_1^\mu q_1^\alpha}{q_1^2}}{q_1^2-m_{X(3872)}^2}  
\frac{1}{q^{2}-m_{D}^{2}}  \frac{1}{q_2^{2}-m_{\eta}^{2}}\\ \nonumber   && [\varepsilon_{\mu}(k_0)][\varepsilon_{\alpha}(p_1)][q_2^\beta\varepsilon_{\beta}(p_2)]F^2,   \\ \nonumber 
 i\mathcal{M}^{c,d}&=&g_{X(4412)X(3872)\eta} g_{X(3872)D\bar{D}^*}g_{D^*D\eta} \int \frac{d^{4}q}{(2\pi)^4} \frac{-g^{\mu\alpha}+\frac{q_1^\mu q_1^\alpha}{q_1^2}}{q_1^2-m_{X(3872)}^2}  
\frac{-g^{\alpha a}+\frac{q^\alpha q^a}{q^2}}{q^{2}-m_{D}^{2}}  \frac{1}{q_2^{2}-m_{\eta}^{2}} \\ \nonumber   && [\varepsilon_{\mu}(k_0)][q_{2a}]F^2,   \\ \nonumber 
 i\mathcal{M}^{e,f}&=&g_{X(4412)X(3872)\eta} g_{X(3872)D\bar{D}^*}g_{D^*D^*\eta}  \int \frac{d^{4}q}{(2\pi)^4} \frac{-g^{\mu\alpha}+\frac{q_1^\mu q_1^\alpha}{q_1^2}}{q_1^2-m_{X(3872)}^2}  
\frac{-g^{\alpha a}+\frac{q^\alpha q^a}{q^2}}{q^{2}-m_{D}^{2}}  \frac{1}{q_2^{2}-m_{\eta}^{2}} \\ \nonumber   && [\varepsilon_{\mu}(k_0)][\varepsilon^{bacd} q_b p_{2c}\varepsilon_{d}(p_2)]F^2,
 \end{eqnarray}
 where $F$ is the form factor.  The form factor not only reduces the divergence of loop integrals but also considers the internal structures of hadron interaction vertices. The form factor is written as $F=\frac{\Lambda^2-m^2}{\Lambda^2-q^2}$, where the cutoff $\Lambda$ is further parameterized as a dimensionless parameter $\alpha$, i.e., $\Lambda=m+\alpha \Lambda_{QCD}$. Following Ref.~\cite{Wu:2025fzx}, the parameter $\alpha$ varies from $1$ to $2$.

Following Ref.~\cite{Wu:2025fzx}, we estimate the production rate of the three-body molecule $X(4412)$ in $B$ meson decays.
The production mechanism is illustrated in Fig.~\ref{productions}. The process proceeds as follows: the $B$ meson first decays into $X(3872)$ and a $K^*$ meson; subsequently, the $K^*$ meson undergoes scattering into a $K$ and an $\eta$ meson; finally, the three-body $\bar{D}^*D\eta$ molecular state is dynamically generated through final-state interactions between the $X(3872)$ and the $\eta$ meson.  
 
 \begin{figure}[!h]
	\centering
	\includegraphics[width=6.0cm]{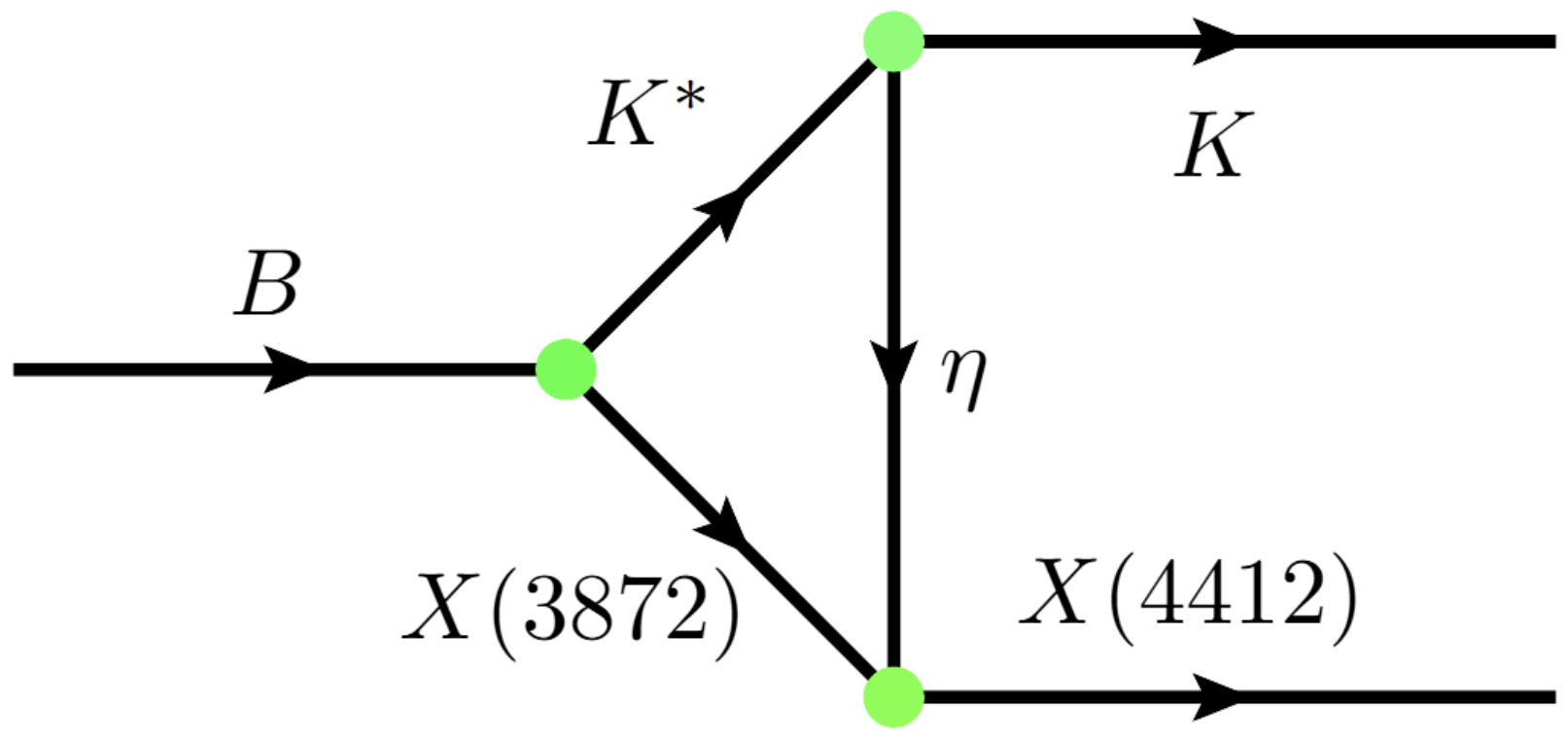}
	\caption{Triangle diagrams for the production of $X(4412)$ in $B$ meson decays.     }\label{productions}
\end{figure}

Similarly, we employ the effective Lagrangian approach to calculate the decay $B \to X(4412) K$.  
The effective Lagrangian describing the interaction vertices of Fig.~\ref{productions} are written as ~\cite{Oh:2000qr,Guo:2013zbw,Dong:2022cuw}
\begin{eqnarray}
\label{dddds}
\mathcal{L}_{K^{\ast} K \eta}&=& -i g_{K^{\ast} K \eta} ( K \partial^{\mu} \eta \bar{K}_{\mu}^{\ast}-  K_{\mu}^{\ast}\partial^{\mu} \eta \bar{K}), \\  
\mathcal{L}_{BX(3872)K^*}&=&  \frac{G_{F}}{\sqrt{2}}V_{cb}V_{cs} a_{2} m_{X(3872)}f_{X(3872)}\varepsilon_{\mu}(q_2)\left[(-g^{\mu \alpha} (m_{X(3872)}+m_{B}) A_{1}\left(q_2^{2}\right)\right.  \\ \nonumber &+& \left. (k_0+q_1)^{\mu} (k_0+q_1)^{\alpha} \frac{A_{2}\left(q_2^{2}\right)}{m_{X(3872)}+m_{B}}  
+i \varepsilon^{\mu \alpha \beta \gamma}(k_0+q_1)_\beta q_{2\gamma} \frac{V\left(q_2^{2}\right)}{m_{X(3872)}+m_{B}}\right]\varepsilon_{\alpha}(q_1),  
\end{eqnarray}
where the couplings are taken as $g_{K^{\ast} K \eta}=5.23$~\cite{Ronchen:2012eg},  $G_F = 1.166 \times 10^{-5}~{\rm GeV}^{-2}$, $V_{cb}=0.041$, $V_{cs}=0.987$, $a_2 = 0.236$, and $f_{X(3872)} = 182$ MeV~\cite{Wu:2023rrp}. The form factors of   $A_{0}(t)$, $A_{1}(t)$, $A_{2}(t)$, and $V(t)$ with $t \equiv q^{2}$ can be parameterized in the following form~\cite{Verma:2011yw} 
\begin{equation}
F(t)=\frac{F(0)}{1-a\left(t / m_{B}^{2}\right)+b\left(t^{2} / m_{ B}^{4}\right)}.
\end{equation}
The values of $F_0$, $a$, and $b$ in  the transition form factors of $B\to K^{\ast}$ are taken from Ref.~\cite{Verma:2011yw} and shown in Table~\ref{BtoKformfactor1}.

\begin{table}[!h]
 \centering
 \caption{Values of  $F(0)$, $a$, $b$ in the $B  \rightarrow K^*$  transition  form factors~\cite{Verma:2011yw}. \label{BtoKformfactor1} }
 \begin{tabular}{c|cccccc}
 \hline\hline
  & $V$~~~ & $A_0$~~~ &  $A_1$~~~ & $A_2$~~~\\
 \hline  
 $F(0)^{B\to K^{(*)}}$~~~  & 0.36~~~ & 0.38~~~ & 0.31~~~ & 0.28~~~\\
 $a^{B\to K^{(*)}}$~~~   & 1.69~~~ & 1.61~~~ & 0.84~~~ & 1.53~~~\\
 $b^{B\to K^{(*)}}$~~~   & 0.95~~~ &0.89~~~ & 0.12~~~ & 0.79~~~\\
  \hline\hline
 \end{tabular}
 \end{table}

With the above Lagrangians, the amplitude of Fig.~\ref{productions} is  written as  
 \begin{eqnarray}  
\label{3872amp}
\mathcal{A}& = &g_{K^*K\eta}g_{X(4412)X(3872)\eta}\int \frac{d^{4} q_{3}}{(2 \pi)^{4}} \frac{\mathcal{A}(B\to X(3872) K^*)^{\mu\alpha}(-g^{\alpha a}+\frac{q_1^{\alpha }q_1^a }{q_1^2})(-g^{\mu u}+\frac{q_2^{\mu }q_2^u }{q_2^2})}{\left(q_{1}^{2}-m_{K^*}^{2}\right)\left(q_{2}^{2}-m_{X(3872)}^{2}\right)\left(q_{3}^{2}-m_{\eta}^{2}\right)}q_{3a}\varepsilon_{u}(p_2)F^2.   \label{38721} 
\end{eqnarray}

  With the  amplitudes of two-body decays, one can calculate  
 the corresponding partial decay widths  as
 \begin{eqnarray}
\Gamma=\frac{1}{2J+1}\frac{1}{8\pi}\frac{|\vec{p}|}{m^2}\bar{|\mathcal{M}|}^{2},
\end{eqnarray}
where $J$ is the total angular momentum of the initial state, the overline indicates the sum over the polarization vectors of final states, and $|\vec{p}|$ is the momentum of either final state in the rest frame of the initial state.

 \begin{figure}[htph]
\begin{center}
\begin{tabular}{ccc}
\subfigure[]
{
\begin{minipage}[t]{0.32\linewidth}
\begin{center}
\begin{overpic}[scale=.24]{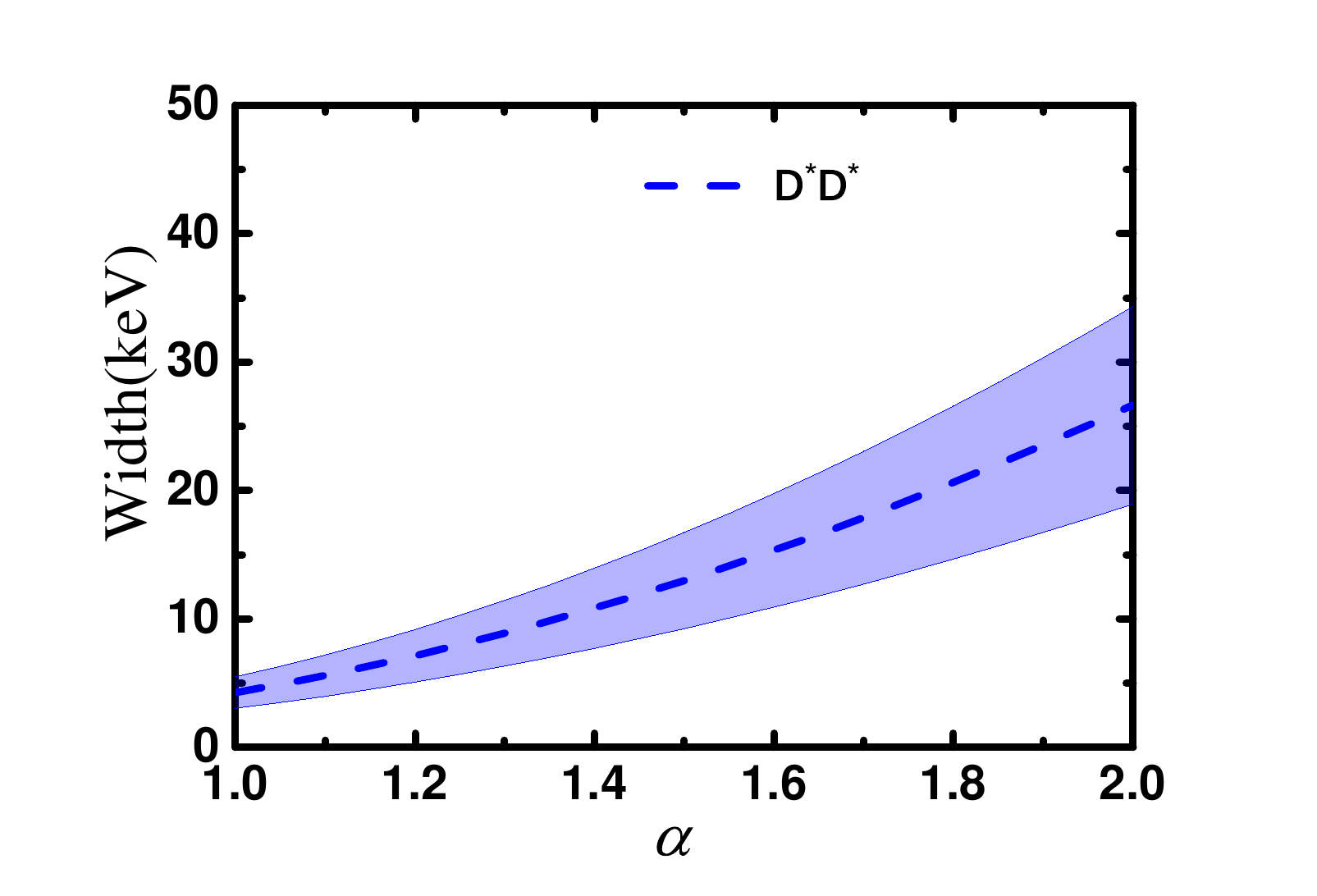}
\end{overpic}
\end{center}
\end{minipage}
}
\subfigure[]
{
\begin{minipage}[t]{0.32\linewidth}
\begin{center}
\begin{overpic}[scale=.24]{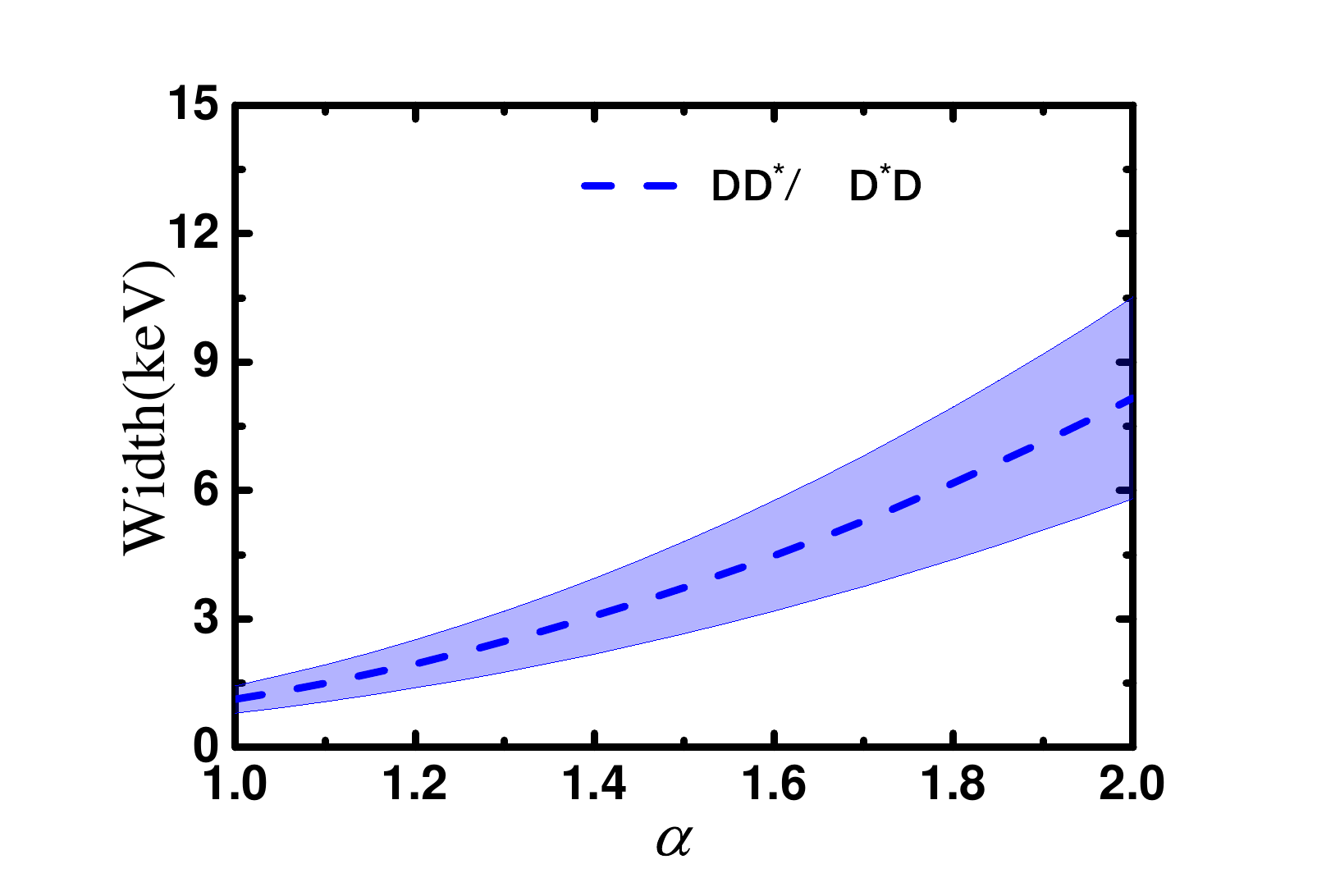}
\end{overpic}
\end{center}
\end{minipage} 
}
\subfigure[]
{
\begin{minipage}[t]{0.32\linewidth}
\begin{center}
\begin{overpic}[scale=.24]{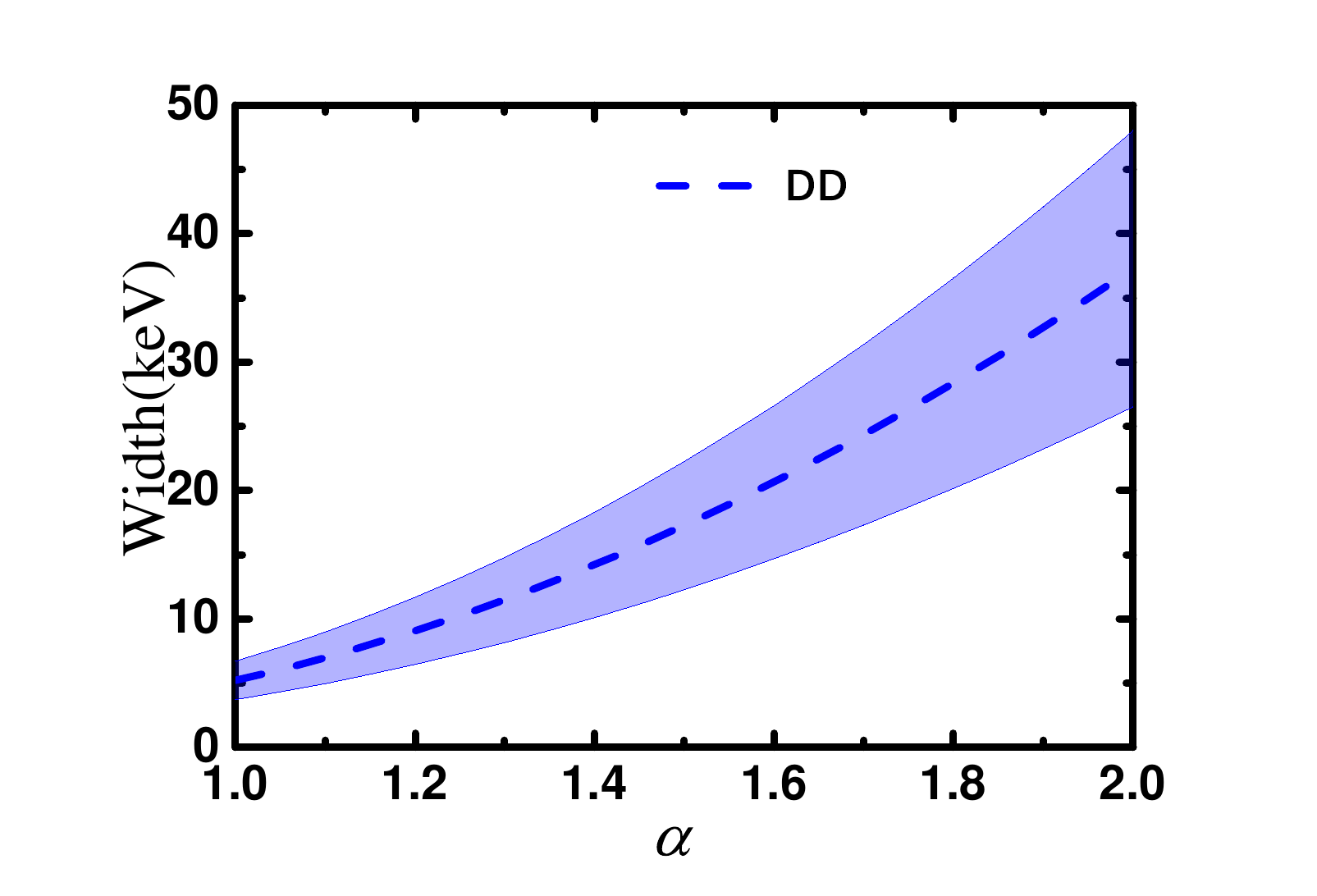}
\end{overpic}
\end{center}
\end{minipage}
}
\end{tabular}
\caption{ Partial decay widths  of $X(4412)\to \bar{D}^*D^*$, $X(4412)\to \bar{D}D^*/\bar{D}^*D$, and $X(4412)\to \bar{D}D$ as functions of $\alpha$. The dashed line and band correspond to the central value and its uncertainty.      }
\label{ytox}
\end{center}
\end{figure}

In our calculations,  the dominant uncertainties originate from the couplings of the three vertices of the triangle diagrams. For the vertices describing the dynamical generation of hadronic molecules, the uncertainties are mainly from the cutoff $\Lambda$ of the form factor~\cite{Wu:2023rrp}. If we increase the cutoff from 1 to 2 GeV, the couplings decrease by approximately 10\%.  Therefore, we assign a $10\%$ uncertainty for the couplings of the molecules to their constituents~\cite{Wu:2023rrp}.   For  the couplings  $g_{D^{(\ast)}D^{(\ast)}\eta}$ and $g_{K^*K\eta}$, 
the   SU(3)-flavor symmetry breaking can lead to an uncertainty of about $20\%$~\cite{FlavourLatticeAveragingGroupFLAG:2021npn}. As for the weak interaction vertices, the experimental uncertainties of the branching fractions of $B$ meson decaying into charmonia and $K^{(*)}$ mesons lead to about $10\%$ uncertainty for the effective Wilson coefficient  $a_{2}$~\cite{Xie:2022lyw}.
 Finally, we obtain the uncertainties of the branching fractions originating from the uncertainties of these parameters via a  Monte Carlo sampling in their 1$\sigma$ intervals.

In Fig.~\ref{ytox}, we present the partial decay widths of the $X(4412)$ as a function of $\alpha$, with the bands reflecting uncertainties arising from the vertex couplings.  The partial widths for $X(4412)\to \bar{D}^*D^*$ and $X(4412) \to \bar{D}D$ are on the order of tens of keV, while those for $X(4412) \to \bar{D}^*D$ and $X(4412) \to \bar{D}D^*$ are around several keV.
It is worth noting that although the absolute partial widths exhibit notable uncertainties, the relative uncertainty in their ratios is greatly reduced because systematic errors largely cancel across these similar decay channels. We find the ratio $\Gamma(X(4412)\to \bar{D}^*D/\bar{D}D^*):\Gamma(X(4412)\to \bar{D}^*D^*):\Gamma(X(4412)\to \bar{D}D)\approx 0.4:0.8:1$. Assuming that the two-body partial decays of the $X(4412)$ play the dominant role, we can roughly estimate the branching fractions of these decays $\mathcal{B}[X(4412)\to \bar{D}^*D/\bar{D}D^*]=18\%$, $\mathcal{B}[X(4412)\to \bar{D}^*D^*]=36\%$, and $\mathcal{B}[X(4412)\to \bar{D}D]=46\%$. In the isospin limit, we obtain the branching fractions of $\mathcal{B}[X(4412)\to \bar{D}^{0}D^{*0}/D^{*+}D^{-}/\bar{D}^{*0}D^0/D^{+}D^{*-}]=4.5\%$,  $\mathcal{B}[X(4412)\to \bar{D}^{*0}D^{*0}/D^{*+}D^{*-}]=18\%$, and $\mathcal{B}[X(4412)\to \bar{D}^0D^0/D^+D^-]=23\%$.

 \begin{figure}[!h]
	\centering
	\includegraphics[width=8.0cm]{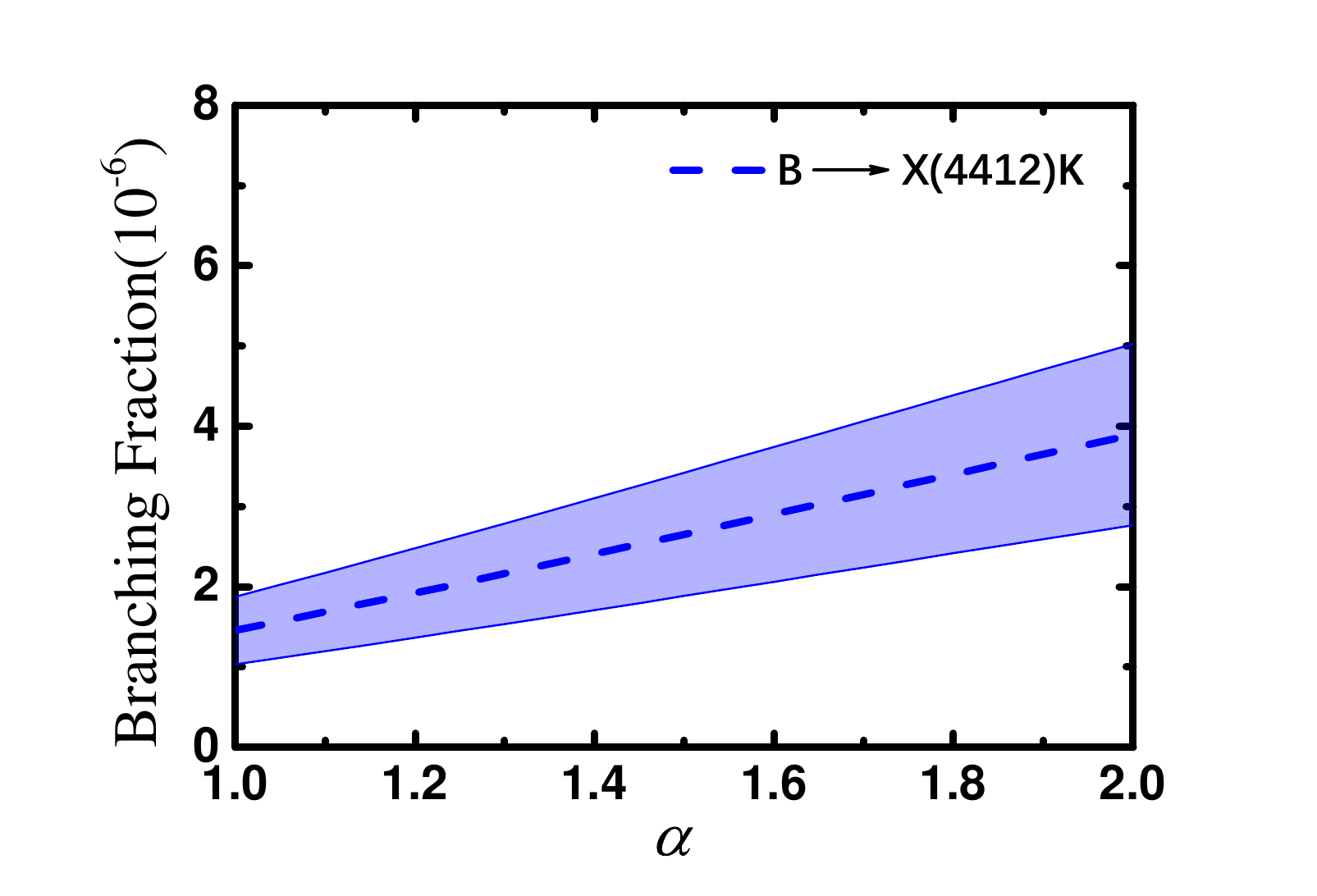}
	\caption{Branching fractions of the decay $B \to X(4412) K$ as a function of $\alpha$.     }\label{resultsp}
\end{figure}

In Fig.~\ref{resultsp}, we show the branching fraction of the decay $B \to K X(4412)$ as a function of $\alpha$.   One can see that the branching fraction of the decay $B \to K X(4412)$ is approximately $2\times 10^{-6}$. Considering the partial decays of $X(4412)$, we obtain the following branching fractions:  $\mathcal{B}[B \to K (X(4412) \to \bar{D}^{0}D^{*0}/D^{*+}D^{-}/\bar{D}^{*0}D^0/D^{+}D^{*-})] \approx 0.1\times 10^{-6}$, $\mathcal{B}[B \to K (X(4412) \to \bar{D}^0D^0/D^+D^- )] \approx 0.5\times 10^{-6}$,  and $\mathcal{B}[B \to K (X(4412) \to \bar{D}^{*0}D^{*0}/D^{*+}D^{*-})] \approx 0.4\times 10^{-6}$.   With the branching fraction $\mathcal{B}(B \to D^{*\pm}D^{\mp} K)=6\times 10^{-4} $~\cite{ParticleDataGroup:2024cfk}, we obtain the ratio  $\mathcal{B}[B \to K (X(4412) \to D^{*\pm}D^{\mp})]/\mathcal{B}(B \to D^{*\pm}D^{\mp} K)\approx 2\times 10^{-4}$.  The LHCb Collaboration reported about $ 2 \times 10^{3}$ events for the decay $B^+ \to D^{*\pm}D^{\mp} K^+$ at an integrated luminosity of $9~\text{fb}^{-1}$~\cite{LHCb:2024vfz}. This event number is expected to reach at least $2$ and $10$ for integrated luminosities of $50~\text{fb}^{-1}$ and $350~\text{fb}^{-1}$, respectively.     
  With $\mathcal{B}(B \to D^{+}D^{-} K) = 2.2 \times 10^{-4}$~\cite{ParticleDataGroup:2024cfk}, the corresponding ratio is  $\mathcal{B}[B \to  (X(4412) \to D^{+}D^{-})K]/\mathcal{B}(B \to D^{+}D^{-} K)\approx 2\times 10^{-3}$.  For the decay $B \to D^{+}D^{-} K$, the LHCb Collaboration observed approximately $1 \times 10^{3}$ events at $9~\text{fb}^{-1}$~\cite{LHCb:2020bls}. These events would increase to at least $10$ and $80$ for luminosities of $50~\text{fb}^{-1}$ and $350~\text{fb}^{-1}$, respectively.  Using $\mathcal{B}(B \to D^{*+}D^{*-} K) = 1.32 \times 10^{-3}$~\cite{ParticleDataGroup:2024cfk}, we obtain the ratio  $\mathcal{B}[B \to  (X(4412) \to D^{*+}D^{*-})K]/\mathcal{B}(B \to D^{*+}D^{*-} K)\approx 3\times 10^{-4}$. The event yield for $B \to D^{*+}D^{*-} K$ was measured to be around only $2 \times 10^{2}$ by the BaBar collaboration~\cite{BaBar:2010tqo}, making the detection of $X(4412)$ challenging in this channel. Therefore, we recommend experimental searches for the $1^{-+}$ $\bar{D}^*D\eta$ molecule in the decay channels $B \to D^{*\pm}D^{\mp} K$ and $B \to D^{+}D^{-} K$.

\end{widetext}

\end{document}